\documentclass{cjour}
\usepackage{epsfig}
\def\simless{\mathbin{\lower 2pt\hbox
   {$\rlap{\raise 5pt\hbox{$\char'074$}}\mathchar"7218$}}}   
\def\simgreat{\mathbin{\lower 2pt\hbox
   {$\rlap{\raise 5pt\hbox{$\char'076$}}\mathchar"7218$}}}   

\authorrunninghead{M. R. Bate}
\titlerunninghead{Recent Advances in Binary Star Formation Using SPH}

\begin{document}

\title{Recent Advances in Binary Star Formation Using SPH}


\author{Matthew R. Bate}
\affil{Institute of Astronomy, Madingley Road, Cambridge CB3 0HA, United Kingdom}
\email{mbate@ast.cam.ac.uk}

\abstract{We review recent advances in the study of binary star formation that have been made using the smoothed particle hydrodynamics technique.}

\begin{article}

\section{Introduction}
 
The Smoothed Particle Hydrodynamics (SPH) numerical method 
was introduced by Lucy \cite{Lucy77} and 
Gingold and Monaghan \cite{GinMon77}.  Its first application
was in the field of star formation, where it was used to 
study whether or not a rapidly-rotating polytrope could undergo
fission to form a close binary system \cite{Lucy77, GinMon78}.
Since this initial application, SPH has been widely used in the
study of star formation, for example \cite{GinMon81, 
MiyHayNar84, Lattanzioetal85, Durisenetal86, SigKla97, 
BonBas91, BatBonPri95, Heller93, NelPap93, ArtLub94,
Larwoodetal96, Nelsonetal98, VanCam98, 
Chapmanetal92}.

SPH is has many attributes which make it particularly 
well suited to the study of star formation.
SPH is Lagrangian and does not require a computational grid.
Thus, it can efficiently follow problems with large density 
contrasts since computational effort is not wasted simulating 
the low-density regions.  Also, recent SPH implementations 
\cite{Evrard88, HerKat89, Benz90, Monaghan92} use spatially 
and temporally varying smoothing lengths so that the resolution 
increases automatically with increasing density; the complex 
multi-grid and adaptive-grid schemes that are used for 
finite difference methods are avoided.

In this proceedings, we review some of the recent advances
in the study of binary star formation that have been made using 
SPH.  In Section \ref{jeansmass}, we discuss the importance of
always resolving the Jeans mass in numerical studies of self-gravitating
gas.  While this has been demonstrated using various types of 
hydrodynamic code, we concentrate specifically on the problems
that can arise if this criterion is not obeyed with SPH.
In Section \ref{collstellar}, we demonstrate that, for the first time, 
it is now possible to perform three-dimensional hydrodynamic 
calculations which follow the collapse of a molecular cloud 
core to stellar densities.  These calculations are performed with
SPH.  Finally, in Section \ref{accretion}, 
we discuss how SPH has been used to study the evolution of a 
protobinary system as it accretes from an infalling gaseous 
envelope and how this work can lead to predictions
of the properties of binary stars.

\begin{figure}[t]
\begin{center}
\epsfig{file=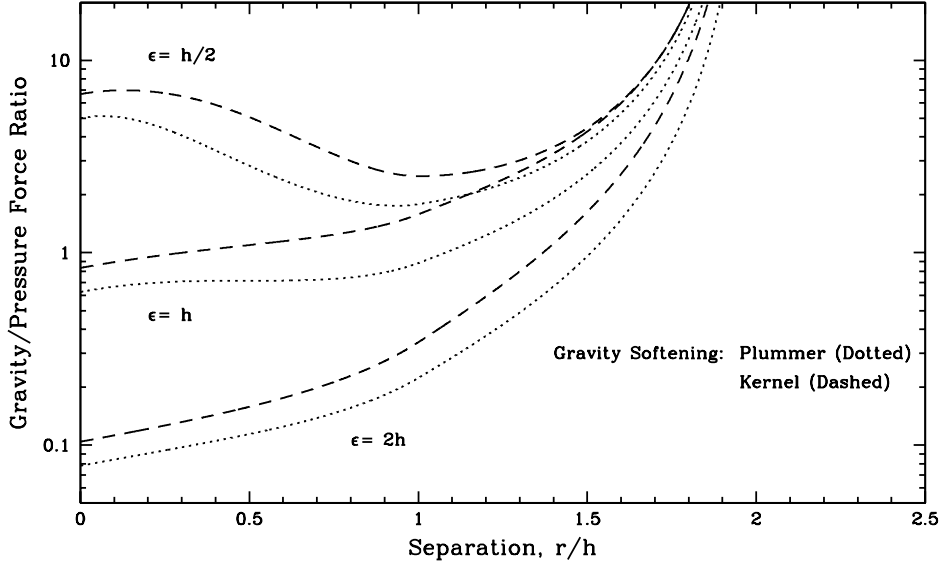, width=13cm}
\caption{\label{bate.forces} The ratio of the gravitational force 
to pressure force between two SPH particles in a Jeans-mass
clump of gas of radius $R=h$.  The gravitational softening length
is $\epsilon$ and the hydrodynamic smoothing length is $h$.}
\end{center}
\end{figure}

\section{The Importance of Resolving the Jeans Mass}
\label{jeansmass}

It has recently been realised that it is important that the Jeans
mass/length is always resolved during a hydrodynamical calculation.  
This has been demonstrated both with SPH \cite{BatBur97, Whitworth98}
and with grid-based codes \cite{Trueloveetal97, Trueloveetal98, Boss98}.  
If this criterion is not obeyed, artificial 
fragmentation can be induced, or fragmentation can be inhibited.  
Essentially this is because when the resolution length/mass 
approaches the Jeans length/mass, collapse is artificially delayed
due to viscous forces, softening of gravitational forces, or a combination
of both.  A good example is the collapse of an isothermal 
filament \cite{Trueloveetal97}.  Such a filament should collapse
without limit to a filamentary singularity without fragmenting.  
However, if the collapse perpendicular to the major-axis is delayed, small
density perturbations along the filament may have enough time to grow 
to non-linear amplitudes and fragments may
form along the bar.

\begin{figure}[t]
\begin{center}
\epsfig{file=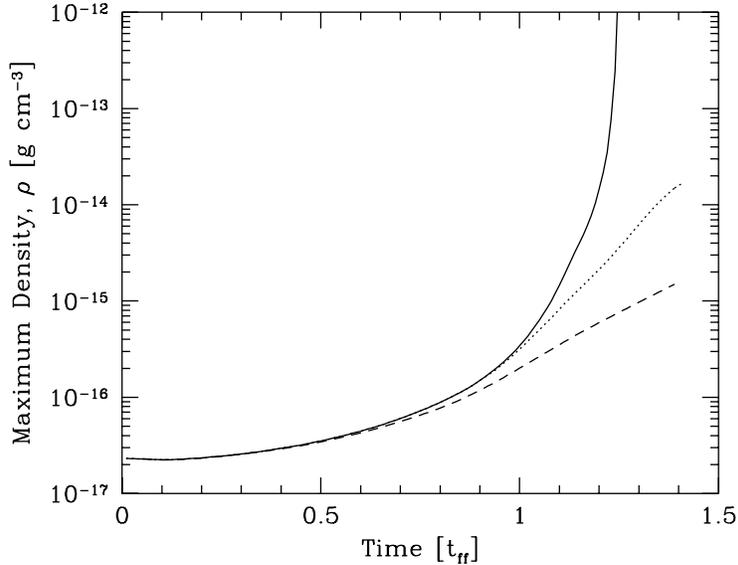, width=10cm}
\caption{\label{bate.massden} Maximum density versus time for during the
collapse of a molecular cloud core [8].  Three calculations were 
performed using SPH with $\epsilon=h$.  The smoothing
lengths were allowed to vary with time and space freely (solid line) or 
subject to
a minimum length of 5\% (dotted line) or 10 \% (dashed line) of the
initial cloud radius.  When the minimum Jeans length in the 
calculation is less than or approximately equal to the smoothing/softening
length, the collapse is delayed.
}
\end{center}
\end{figure}

Bate \& Burkert \cite{BatBur97} first demonstrated the need for the
Jeans mass criterion using self-gravitating SPH calculations.  With
SPH, the problem can be understood by considering the 
gravitational and pressure forces between two particles 
within a marginally Jeans-stable clump of gas of radius $R\approx h$
(Figure \ref{bate.forces}).  SPH codes typically soften 
the gravitational forces between neighbouring particles, using either 
the Plummer force law or kernel softening 
\cite{GinMon77, HerKat89, Benz90}.  The
characteristic gravitational softening length, $\epsilon$, 
may or may not be equal to the SPH hydrodynamic smoothing 
length, $h$, depending on the specific implementation.
If $\epsilon=h$, the ratio of the gravitational and 
pressure forces between two particles is approximately 
constant for particle with separations $\simless h$.
Thus, a Jeans-unstable clump of gas will collapse, while a 
Jeans-stable clump will be supported.  However, although the
ratio of the gravitational and pressure forces is approximately 
independent of the separation of the particles, the magnitude 
of the gravitational force decreases with separation due to 
the softening.  Thus, while a Jeans-unstable clump of gas with 
a size much larger than the resolution length, $h$, 
will collapse at the correct rate, the collapse of clumps with 
a size $\approx h$ will be delayed.  This is demonstrated
in Figure \ref{bate.massden}.  

\begin{figure}[t]
\begin{center}
\epsfig{file=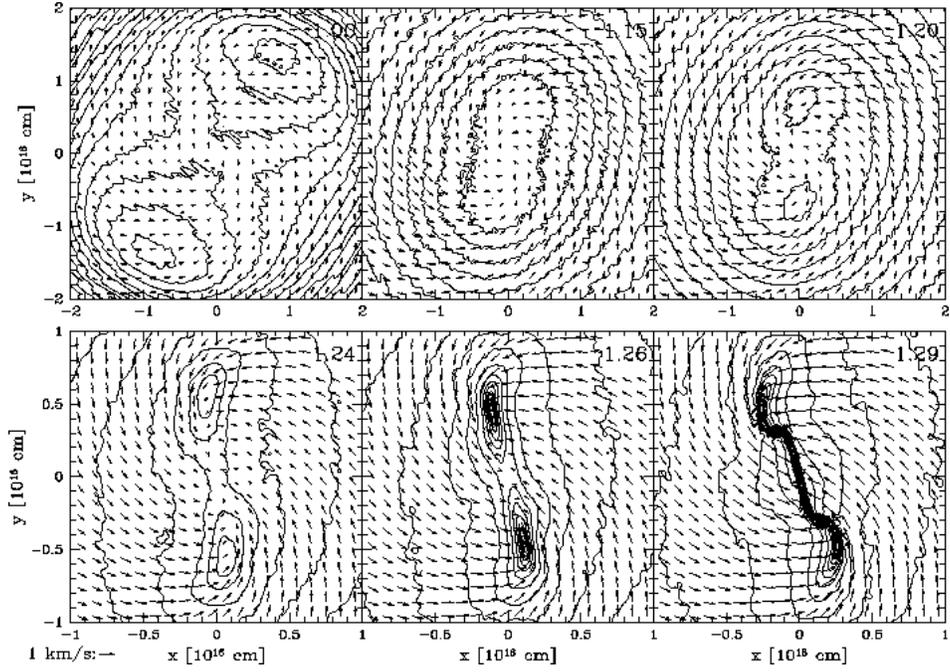, width=12.5cm}
\vspace{0.25truecm}
\caption{\label{bate.80000} Collapse and fragmentation of a 
molecular cloud core to form a binary [8]. The initial cloud had an initial
10\% $m=2$ density perturbation.  The local Jeans mass/length is unresolved
inside the thick density contour.  The calculation was performed
with $8\times 10^4$ particles.
}
\vspace{-0.5truecm}
\end{center}
\end{figure}

Unfortunately, the effect is not always limited to a simple delay of 
the collapse.  Given the right problem, artificial fragmentation
can be induced (such as with the filament described above), 
or inhibited.  In Figure \ref{bate.80000} we show how the collapse
of a particular molecular cloud core with an initial $m=2$ 
density perturbation results in a binary protostellar system
with a bar of gas between them.  This calculation was performed
with $8\times 10^4$ particles, enough to resolved the Jeans mass/length
until after the binary had formed.  However, performing
the same calculation with $1\times 10^4$ particles gives a different
result: a single, dense bar of gas without the binary.  In this 
calculation, the Jeans mass becomes unresolved before $t=1.20$
and the collapse of each of the two over-dense regions resulting from the 
original $m=2$ density perturbation is delayed.  The collapse of the
larger-scale elongated region, however, continues, leading to 
the formation of a bar rather than a binary.

\begin{figure}[t]
\begin{center}
\epsfig{file=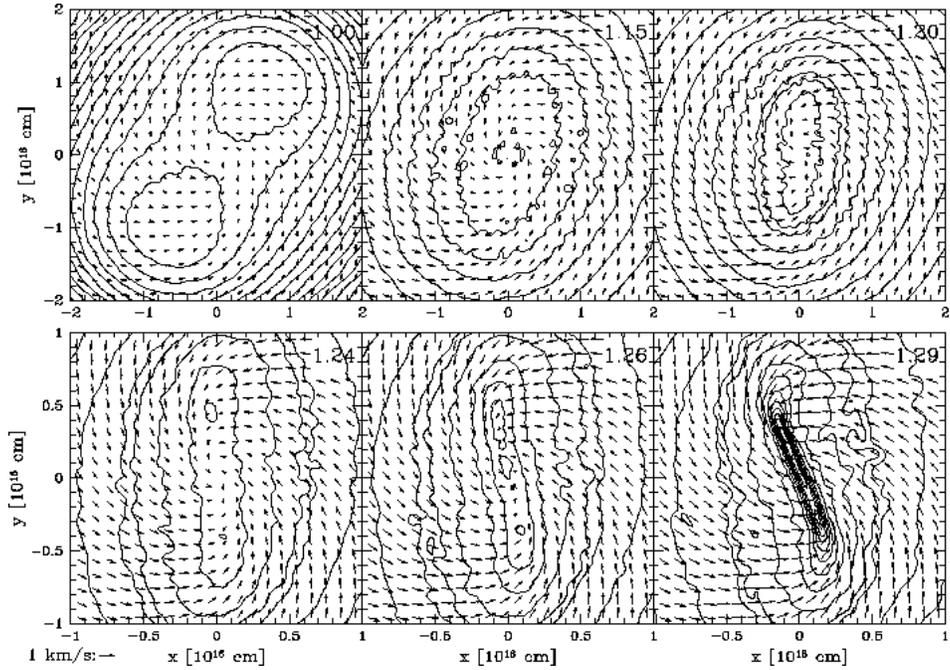, width=12.5cm}
\vspace{0.25truecm}
\caption{\label{bate.10000} Same as Figure 3, except that the calculation
was performed with $1\times 10^4$ particles.  When $t \simgreat 1.20$,
the region within which the binary formed in the $8\times 10^4$-particle
calculation becomes unresolved, collapse of the two over-dense
regions is delayed, and the calculation eventually produces a dense
bar instead of a binary.
}
\vspace{-1.0truecm}
\end{center}
\end{figure}

\begin{figure}[t]
\begin{center}
\epsfig{file=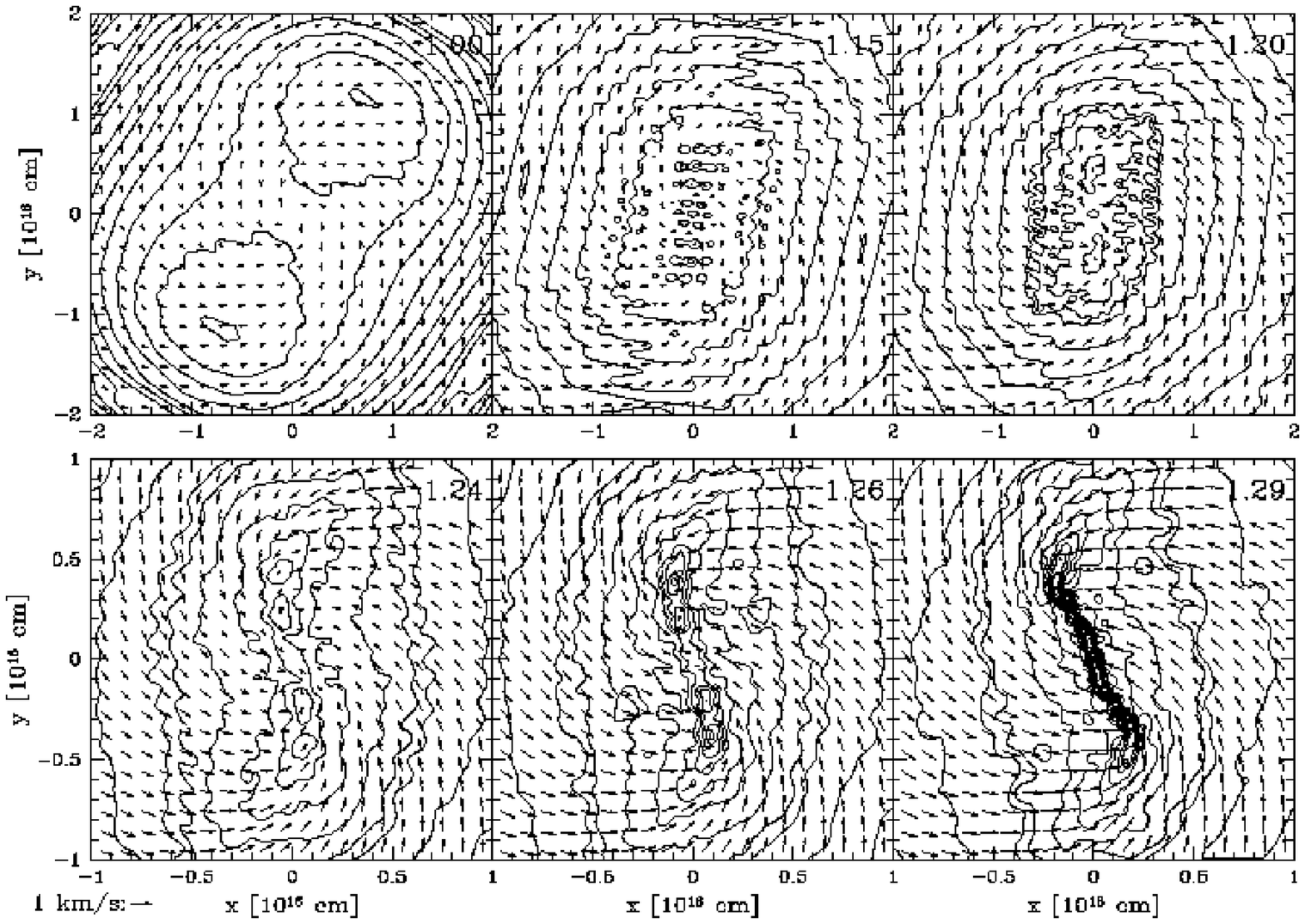, width=12.5cm}
\vspace{0.25truecm}
\caption{\label{bate.10000eh} Same as Figure 4, except that the calculation
was performed with $\epsilon=1.0\times 10^{14}$ cm (i.e.~$e<h$) 
instead of $e=h$.
When the local Jeans-mass becomes unresolved ($t \simgreat 1.20$),
the calculation becomes susceptible to artificial fragmentation.
Instead of a binary being formed, each of the two over-dense
regions undergoes binary fragmentation 
so that the final result is a quadruple system.
}
\end{center}
\end{figure}

It is important to note that, with SPH, the effect of not 
resolving a Jeans mass also depends on the way the
gravitational softening and hydrodynamic smoothing are implemented!
If $\epsilon < h$, the ratio of the gravitational to pressure forces
between two particles increases with decreasing separation 
(Figure \ref{bate.forces}).  This
may lead to an instability in which a group of particles within
a Jeans-stable clump collapse artificially.  As demonstrated in
Figure \ref{bate.10000eh}, this can lead to artificial 
fragmentation.  Alternately, if $\epsilon > h$, the pressure forces
between particles within a Jeans-unstable clump may exceed the 
gravitational forces and the clump will be artificially supported
against collapse.

Clearly, the best SPH implementation is one where $\epsilon=h$ always.
In this case, collapse of Jeans-unstable clumps with a size 
similar to that of the resolution length will still be 
delayed, but the possibilities of
artificial collapse within Jeans-stable clumps or the supporting
of Jeans-unstable clumps against collapse are eliminated.  Then,
in order to avoid the collapse of Jeans-unstable clumps being
delayed significantly, enough particles should be used so that
the Jeans length/mass is always resolved.  

How many particles
are necessary to ensure that the Jeans length/mass is resolved?
With SPH, the spatial resolution 
is given by the smoothing length which is usually variable in 
time and space.  The smoothing lengths are set by ensuring that 
each particle contains a certain number of neighbours, 
$N_{\rm neigh}$, or equivalently a fixed mass, within two 
smoothing lengths.  Thus, in contrast to a grid-based code 
which has spatially-limited resolution, SPH has mass-limited 
resolution which {\it automatically} gives greater spatial 
resolution in regions of higher density.  Therefore, with SPH,
it is necessary to ensure that the minimum resolvable mass is 
always less than the Jeans mass.  In practice, Bate \& Burkert 
found that a Jeans mass should always 
be represented by at least $\approx 2 N_{\rm neigh}$ particles. 
\newpage

\section{Collapse of a Molecular Cloud to Stellar Densities}
\label{collstellar}

\begin{figure}[t]
\begin{center}
\epsfig{file=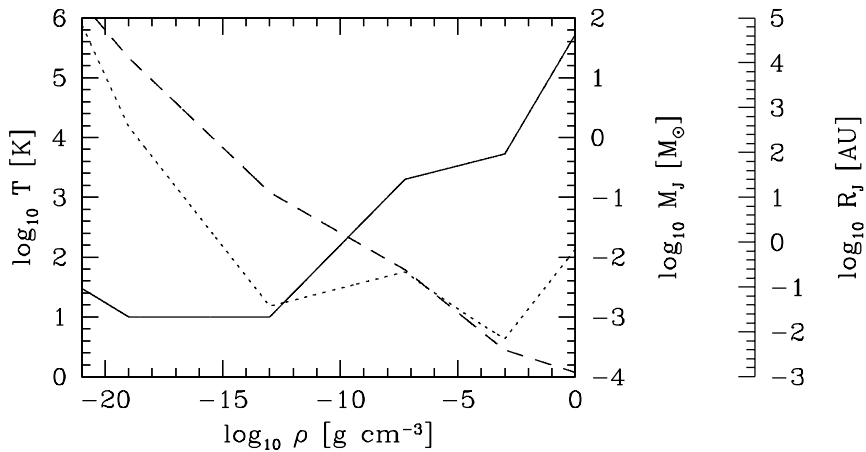, width=12cm}
\caption{\label{bate.tmr} The gas temperature $T$ (solid line), Jeans 
mass $M_{\rm J}$ (dotted line), and Jeans radius $R_{\rm J}$ (dashed 
line), as a function of density in a collapsing molecular cloud core 
(adapted from Tohline [36]).  The minimum Jeans mass of 
$ \approx 4 \times 10^{-4}~{\rm M}_{\odot}$ occurs 
at a density of $\sim 10^{-3} {\rm \ g \ cm}^{-3}$.}
\end{center}
\end{figure}

The mass-limited resolution of SPH is ideal for studying the 
collapse and fragmentation of molecular cloud cores because 
there is a minimum Jeans mass in the problem 
(Figure \ref{bate.tmr}).  By contrast,
there is no minimum Jeans length.  This is a problem 
for grid-based codes which must resort to nested or 
adaptive grids \cite{ BurBod93, Trueloveetal97, Trueloveetal98}.  
With SPH, if the number of 
particles used is sufficient to resolve the minimum Jeans mass, a 
calculation can be followed to arbitrary densities with the required
spatial resolution given automatically with increasing density.

Recently, this ability of SPH was used to perform the first 
three-dimensional calculations ever to follow the collapse of
a molecular cloud core to stellar densities \cite{Bate98}.
The calculations followed the collapse of an initially 
uniform-density molecular cloud core of mass $M=1 {\rm \ M}_\odot$ and
radius $R=7 \times 10^{16} {\rm \ cm}$.

The minimum resolvable mass in the SPH code was 
$\approx 2 N_{\rm neigh} = 100$ particles. Thus, to enable the
minimum Jeans mass during the calculation
($\approx 4 \times 10^{-4}\ {\rm M}_{\odot}$)
to be resolved, the calculation used $3 \times 10^5$ 
equal-mass particles.

The code did not include radiative transfer.  Instead, to model 
the behaviour of the gas during the different phases of collapse,
a piece-wise polytropic equation of state, $P=K \rho^{\gamma}$, was
used, where $P$ is the pressure, $\rho$ is the density, $K$ 
gives the entropy of the gas, and the ratio of specific heats, 
$\gamma$, was varied as
\begin{equation}
\label{polytropic}
\gamma = \cases {\begin{array}{lll}
1       &                         & \rho \leq 1.0 \times 10^{-13} \cr
7/5     & \ 1.0 \times 10^{-13}  < \hspace{-6pt} & \rho \leq 5.7 \times 10^{-8} 
\cr
1.15    & \ 5.7 \times 10^{-8\ } < \hspace{-6pt} & \rho < 1.0 \times 10^{-3} \cr
5/3     &                         & \rho > 1.0 \times 10^{-3} \cr
\end{array}}
\end{equation}
where the densities are in ${\rm g\ cm}^{-3}$ 
(see Figure \ref{bate.tmr}).
The values of $\gamma$ and the transition densities 
were derived from Tohline \cite{Tohline82}.  The variable value of $\gamma$ 
mimics the following behaviour of the gas.  The
collapse is isothermal ($\gamma=1$) until the gas becomes optically thick 
to infrared radiation 
at $\rho \approx 10^{-13} {\rm \ g \ cm}^{-3}$, beyond which $\gamma=7/5$
(appropriate for a diatomic gas).
When the gas reaches a temperature of $\approx$ 2000 K 
($\rho = 5.7 \times 10^{-8} {\rm \ g \ cm}^{-3}$), molecular 
hydrogen begins to dissociate and the temperature only slowly increases
with density.  In this phase $\gamma = 1.15$ is used to 
model {\it both} the decreasing mean molecular weight and the slow increase 
of temperature with density, the latter of which has an effective 
$\gamma \approx 1.10$.
Finally, when the gas has fully dissociated
($\rho \approx 10^{-3} {\rm \ g \ cm}^{-3}$), the gas is monatomic and
$\gamma=5/3$.  The value of $K$ is
defined such that when the gas is isothermal, $K = c_{\rm s}^2$
with $c_{\rm s} = 2.0 \times 10^4 {\rm \ cm\ s^{-1}}$, and when $\gamma$
changes the pressure is continuous.

\begin{figure}[t]
\begin{center}
\epsfig{file=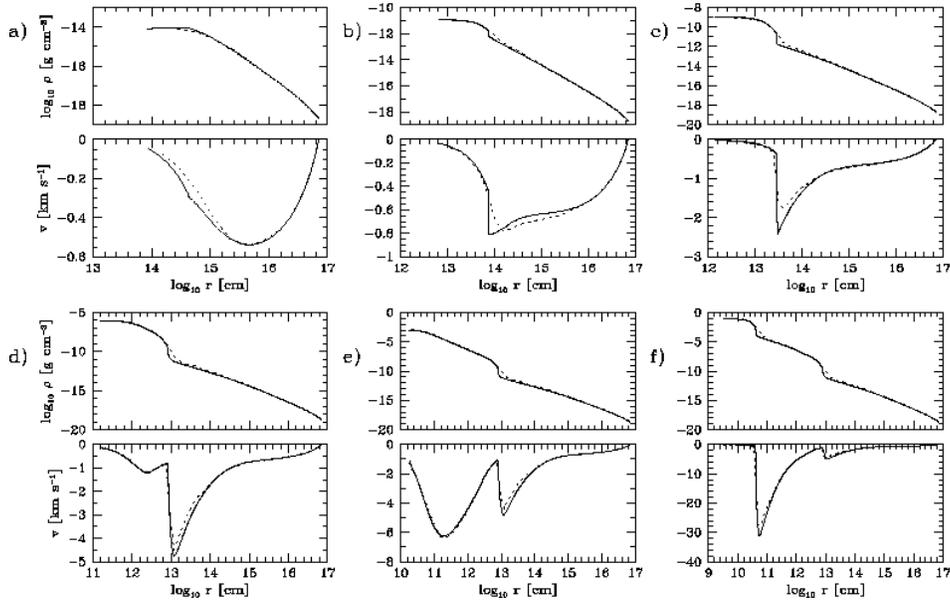, width=12.5cm}
\caption{\label{bate.static}  Radial density and velocity profiles of 
the collapsing molecular cloud core with the one-dimensional grid 
code (solid) and the three-dimensional SPH code (dotted).  
The profiles are compared when the central density in each calculation 
is a) $10^{-14}$, b) $10^{-11}$, c) $10^{-9}$, d) $10^{-6}$, 
e) $10^{-3}$, and f) $10^{-1}$ g cm$^{-3}$.}
\end{center}
\end{figure}

\subsection{Collapse of an initially-static cloud}

To test that the above equation of state captures the important elements
of the gas's behaviour, spherically-symmetric,
one-dimensional (1-D), finite-difference calculations were performed
of the collapse of an initially-static molecular cloud core 
with the above parameters and equation of state.
The results are shown in Figure \ref{bate.static} (solid line).  
These results 
are in good agreement with the results from 1-D calculations 
incorporating radiative transfer 
(e.g.\ Larson 1969; Winkler \& Newman 1980a, b).

The three-dimensional (3-D) SPH code was also tested on the same 
problem to check that the SPH code can indeed accurately resolve the 
collapse down to stellar densities 
(Figure \ref{bate.static}, dotted line).  
There is excellent agreement between the results from the 1-D 
finite-difference code and those from the 3-D SPH code.

\subsection{Collapse of a rotating cloud}

\begin{figure}[t]
\begin{center}
\epsfig{file=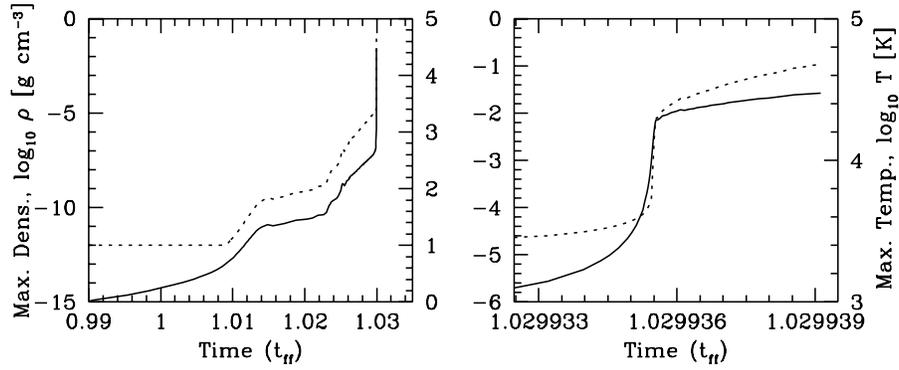, width=12cm}
\caption{\label{bate.rotmassdens} Maximum density (solid line) 
and maximum temperature (dotted line) versus time for the collapsing 
molecular cloud core.  Time is given in units of the initial free-fall 
time ($t_{\rm ff}=1.785 \times 10^{12}$ s).  The right graph has 
expanded axes to show the second collapse phase in greater detail.}
\end{center}
\end{figure}

\begin{figure}[t]
\begin{center}
\epsfig{file=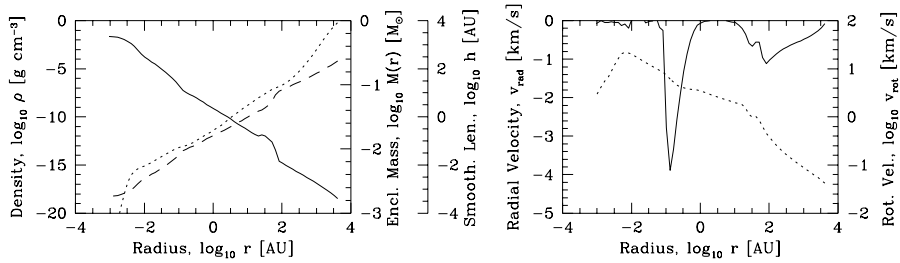, width=12cm}
\caption{\label{bate.finalstate} The state of the system at the end of 
the calculation.  The left graph gives the density (solid line) and 
smoothing length (dashed line) as a function of radius and the mass 
enclosed within radius $r$ of the center of the stellar core (dotted line).  
The right graph gives the radial (solid line) and rotational (dotted line) 
velocities as functions of radius from the center of the stellar core.  
The densities, smoothing lengths and velocities are the mean values in 
the plane perpendicular to the rotation axis and through the stellar core.  
The stellar core ($r<0.004$ AU), inner circumstellar disc 
($0.004<r<0.1$ AU), region undergoing the second collapse ($0.1<r<1$ AU), 
outer circumstellar disc formed from the first core ($1<r<60$ AU), 
and isothermal collapse region ($r>60$ AU) are clearly visible.}
\end{center}
\end{figure}

Three-dimensional calculations are required if the molecular
cloud core is rotating.  In Figures \ref{bate.rotmassdens} to
\ref{bate.finalstate2} we present results from the collapse
of a cloud core which is identical to that in the previous 
section, but which is initially in solid-body rotation with
angular frequency $\Omega=7.6 \times 10^{-14}$ rad s$^{-1}$.
Thus, the ratio of rotational energy to the magnitude of the 
gravitational potential energy is $\beta=0.005$ (i.e.~the cloud
is rotating quite slowly).

\begin{figure}[b]
\begin{center}
\vspace{8.4truecm}
\caption{\label{bate.bar} A time sequence showing the density in 
the plane perpendicular to the rotation axis during the dynamic, 
rotational, bar instability of the first hydrostatic core.  The
panels cover the period from $t\approx 1.023-1.030$ t$_{\rm ff}$.
See also the MPEG animation on this CD-ROM: bate1.mpg.
}
\end{center}
\end{figure}

The evolution of the calculation is as follows 
(Figure \ref{bate.rotmassdens}).  
The initial collapse is isothermal.  When the density surpasses
$10^{-13}$ g cm$^{-3}$, the gas in the center 
is assumed to become optically thick
to infrared radiation and begins to heat ($t=1.009~t_{\rm ff}$).
The heating stops the collapse at the center and the first hydrostatic core
is formed ($t=1.015~t_{\rm ff}$) with maximum density 
$\approx  2 \times 10^{-11}$ g cm$^{-3}$, mass  
$\approx 0.01 {\rm \ M}_\odot$ ($\approx 3 \times 10^3$ particles), and radius 
$\approx 7$ AU.  As the first core accretes, its maximum
density only slowly increases at first.
However, the first core is rapidly rotating, oblate, and has 
$\beta \approx 0.34 > 0.274$, making it dynamically unstable to the
growth of non-axisymmetric perturbations 
\cite{Durisenetal86, ImaDurPic99}. 
At $t \approx 1.023~t_{\rm ff}$, after about 3 rotations, 
the first core becomes violently bar-unstable
and forms trailing spiral arms (Figure \ref{bate.bar}).  This leads
to a rapid increase in maximum density (Figure \ref{bate.rotmassdens})
as angular momentum is removed from
the central regions of the first core (now a disc with spiral structure) 
by gravitational torques ($t=1.023-1.030~t_{\rm ff}$).  An MPEG animation of 
this bar instability is provided on this CD-ROM (bate1.mpg).
When the maximum temperature reaches 2000 K, molecular hydrogen begins to
dissociate, resulting in a rapid second collapse to stellar densities
($t=1.030~t_{\rm ff}$).  The collapse is again halted at a density of 
$\approx 0.007$ g cm$^{-3}$ with the formation of the second hydrostatic, 
or stellar, core.  The initial mass and radius of the stellar core are 
$\approx 0.0015 {\rm \ M}_\odot$ ($\approx 5 \times 10^2$ particles) 
and $\approx 0.8 {\rm \ R}_\odot$, 
respectively.   Finally,
an inner circumstellar disc begins to form around the stellar 
object, within the region undergoing second collapse.  The 
calculation is stopped when the stellar object has a mass of 
$\approx 0.004 {\rm \ M}_\odot$ ($\approx 1.2 \times 10^3$ particles), 
the inner circumstellar disc has extended out to $\approx 0.1$ AU, 
and the outer disc (the remnant of the first hydrostatic core) contains 
$\approx 0.08 {\rm \ M}_\odot$ ($\approx 2.4 \times 10^4$ particles) 
and extends out to $\approx 70$ AU.  Note that the massive outer 
disc forms {\it before} the stellar core.
This final state is illustrated in Figures \ref{bate.finalstate}
and \ref{bate.finalstate2}.
If the calculation was evolved further, the inner circumstellar disc would 
continue to grow in radius as gas with higher angular momentum fell in.
Eventually, the inner disc would meet the outer disc with only a
small molecular dissociation region between the two.

\begin{figure}[t]
\begin{center}
\vspace{18.0truecm}
\caption{\label{bate.finalstate2} The state of the system at the end of 
the calculation (not a time sequence).  
The upper six panels give the density in the plane 
perpendicular to the rotation axis and through the stellar 
core.  The lower six panels give the density in a 
section down the rotation axis.  In each case, the six consecutive panels 
give the structure on a spatial scale that is 10 times smaller than the 
previous panel to resolve structure from 3000 AU to 
$\approx 0.2~{\rm R}_\odot$.  The remnant of the first hydrostatic 
core (now a disc with spiral structure), the inner circumstellar disc, 
and the stellar core are all clearly visible.  The logarithm of the 
density is plotted with the maximum and minimum 
densities (in g~cm$^{-3}$) given under each panel. }
\vspace{-1.0truecm}
\end{center}
\end{figure}

\subsection{Close binary stellar systems}

The ability to perform three-dimensional calculations which follow
the collapse of molecular cloud cores to stellar densities allows
us to study the formation of close ($\simless 1$ AU) binary
stellar systems.  Currently, there is no accepted mechanism for 
forming close binaries; the proposal that close binary systems form
via the fission of a rapidly-rotating protostellar object 
has been discredited by studies of rapidly-rotating 
polytropes \cite{Durisenetal86}.  Although fission
itself appears not to operate, it is possible that fragmentation
can still occur due to the growth of non-axisymmetric perturbations
in rotationally-supported objects.  Only two studies have 
looked at this possibility in detail \cite{Boss87, BonBat94b}, and the
latter of these finds that fragmentation of a massive circumstellar
disc on scales ($\simless 1$ AU) may be possible.  However, in both
these studies, only the region inside the first hydrostatic
core was modelled and the initial conditions were chosen somewhat
artificially.  The ability to perform three-dimensional 
calculations which follow the collapse of molecular cloud 
cores to stellar densities now allows us to study the formation of
close binaries from the collapse of larger-scale ($\approx 10000$ AU)
molecular cloud cores.

\section{The evolution of an accreting protobinary system}
\label{accretion}

The favoured mechanism for the formation of most binary stellar
systems is the fragmentation of a collapsing molecular cloud core.
Fragmentation has been studied numerically for $\approx 20$ years.
These calculations appear to show that it is possible to form binaries 
with similar properties to those that are observed via fragmentation.  
However, they have not allowed us to predict the fundamental properties of 
stellar systems such as the fraction of stellar systems which 
are binary or the properties of binary 
systems (e.g.~the distributions of mass ratios,
separations, and eccentricities and the properties of 
discs in pre-main-sequence systems).

There are two primary reasons for this lack of predictive power.  
First, the results of fragmentation calculations depend sensitively
on the initial conditions, which are poorly constrained.
The second problem is that of accretion.  Fragmentation 
calculations are typically stopped soon after the fragmentation
occurs, when the binary or multiple protostellar system contains 
only a small fraction of the total mass of the
original cloud \cite{Boss86, BonBat94b}.  However, because much
of the gas contained in the original cloud still has to fall on 
to the system and be accreted, 
the final properties of the stellar system are 
unknown.  Following the calculation significantly beyond the
point at which fragmentation occurs is extremely computationally 
intensive.  Thus, it is impossible to perform the number of 
calculations that are required to predict the statistical properties of 
binary stellar systems -- even if we knew the distribution of the 
initial conditions.  On the other hand, if we can overcome this second
difficulty, we can make theoretical predictions about the properties
of binary stars and, by comparing these predictions to the observed
properties of binary systems, we may be able to better constrain
the initial conditions for star formation.

\subsection{The Effects of Accretion on a Protobinary System}

Using SPH, Bate \& Bonnell \cite{BatBon97} studied and quantified how 
the properties of a binary system are affected by the accretion 
of a small amount of gas from an infalling gaseous envelope.  
They found that the effects depend primarily on the specific 
angular momentum of the gas and the binary's mass ratio 
(see also \cite{Artymowicz83, Bate97}).  Generally, accretion of gas 
with low specific angular momentum decreases the mass ratio and 
separation of the binary, while accretion of gas with high 
specific angular momentum increases the separation and drives the 
mass ratio toward unity.  From these results, they predicted 
that closer binaries should have mass ratios that are biased 
toward equal masses compared to wider systems, since the gas which
falls on to a closer system is likely to have more specific angular
momentum, relative to the binary, than for a wider system.

\begin{figure}[t]
\begin{center}
\vspace{9truecm}
\caption{\label{bate.accretion} The dependence of the 
distribution of gas around an accreting protobinary system
on the specific angular momentum of the infalling gas 
(increasing from left-to-right and downward).  For gas with low
specific angular momentum, only the primary forms a circumstellar
disc and the secondary accretes a small amount of gas via a
Bondi-Hoyle-type accretion stream.  For gas with intermediate angular 
momentum, both the primary and secondary form circumstellar discs.
For gas with high angular momentum, two circumstellar discs and
a circumbinary disc are formed.  Finally, for the case with
the highest angular momentum, all the infalling gas settles into
a circumbinary disc.  The
binary has a mass ratio of $q=0.6$ and the primary is on 
the right.
}
\end{center}
\end{figure}

They also studied the process of disc formation around an 
accreting protobinary system and found that for each protostar,
a circumstellar disc was only formed if the specific angular 
momentum of the infalling gas was greater than the specific
orbital angular momentum of that protostar about the centre of
mass of the binary (Figure \ref{bate.accretion}).  This is because,
to be capture by one of the protostars, the gas much achieve the
same specific orbital angular momentum as that of the protostar.
If the gas has more specific angular momentum initially, some of its
angular momentum goes into forming a disc around the protostar.
However, if it has less specific angular momentum initially, there is
no excess angular momentum to form a circumstellar
disc, and it must gain angular momentum even to be captured by
the protostar.  In this case, the infalling gas gains angular 
momentum as it falls on to the protostar in a Bondi-Hoyle-type accretion
stream.  In practice, 
this means that a circumstellar disc is almost always formed around
the primary, but the secondary does not have a circumstellar disc
unless the infalling gas has more specific angular momentum that
some critical value.  In a similar way, the formation of a circumbinary 
disc only begins when the specific angular momentum of the infalling
gas is great enough for the gas to form a circular orbit at a radius
greater than that of the secondary from the centre of mass of the
binary.

\subsection{Development of a Protobinary Evolution Code}

Using the quantitative results of Bate \& Bonnell \cite{BatBon97}, 
Bate \cite{Bate2000} developed a protobinary evolution (PBE) code 
which follows the evolution of a protobinary system as it accretes 
from its initial to its final mass, but does so in far less time 
than would be required for a full hydrodynamic calculation. 

\begin{figure}[t]
\begin{center}
\epsfig{file=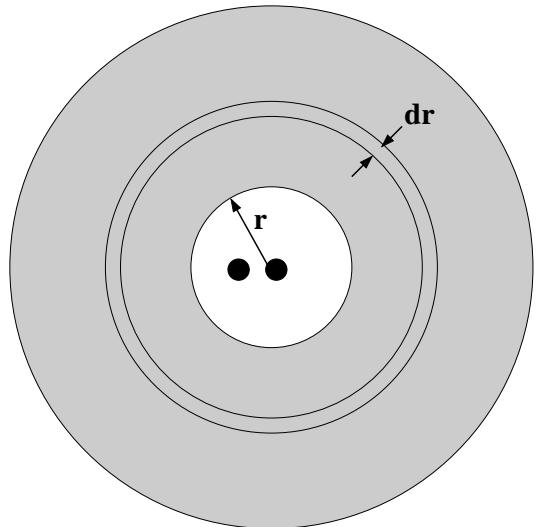, width=7cm}
\caption{\label{bate.model} The model for studying the evolution of a 
protobinary system that forms within a collapsing molecular cloud 
core and is built up to its final mass by accreting the remaining 
gas as it falls on to the binary.}
\end{center}
\end{figure}

This code is based on the following model for the 
formation of binary stellar systems (Figure \ref{bate.model}).  
The model begins with a 
molecular cloud core of known initial density and angular 
momentum profile.  It is assumed that this cloud begins to collapse
and that a `seed' binary system
is formed at the centre, presumably 
via some sort of fragmentation.  The `seed' binary has mass 
ratio $q \leq 1$,
separation $a$, and is assumed to have a circular orbit.
It initial consists of only a small fraction of the total mass of the core
and is assumed to have formed from the gas that was originally
contained within a sphere of radius $r$, at the centre of the initial 
cloud (Figure \ref{bate.model}).  For the results presented in
this proceedings, the separation of the `seed' binary is set by assuming
that the angular momentum of the gas from which the binary 
forms is equal to the orbital angular momentum of the binary.

Subsequently, the binary accretes the remainder of the initial 
cloud (which falls on to the binary) and the binary's 
properties evolve due to the accretion.  This evolution is calculated
by taking a thin shell of gas of thickness d$r$ 
(Figure \ref{bate.model}), surrounding the
sphere from which the binary was formed, dividing the shell into 
small elements of gas, and calculating the effect that each 
element of gas has on the protobinary when it is accreted
(using the results of Bate \& Bonnell \cite{BatBon97}).  
The binary's parameters (masses and separation) are updated,
and the next shell of gas is considered until the whole cloud 
is accreted on to the binary.  The amount of gas which settles 
into a circumbinary disc is also recorded.  In this way, the code
calculates the 
evolution of the binary from its initial to its final state when
all of the original cloud's gas is contained
either in one of the two stars or their surrounding discs.

\subsection{Testing the Protobinary Evolution Code}

\begin{figure}[t]
\begin{center}
\epsfig{file=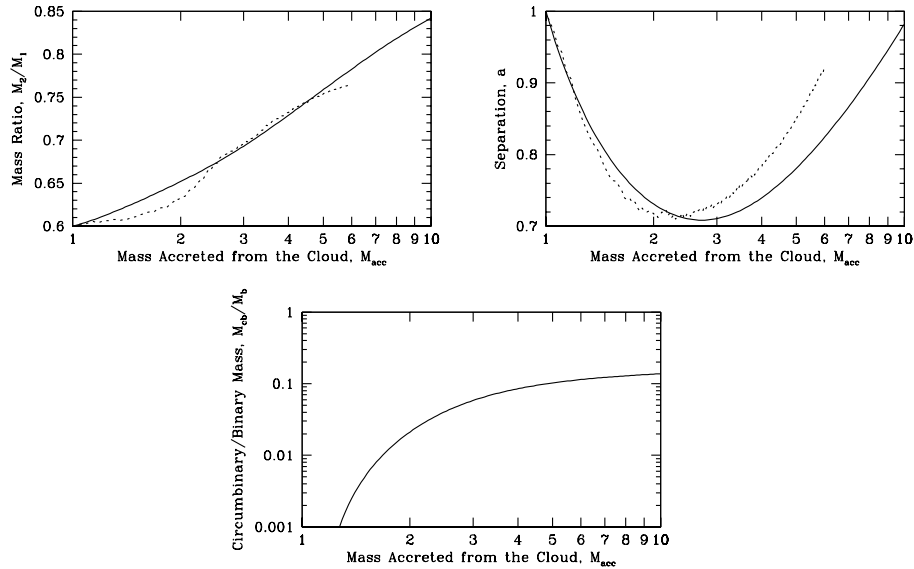, width=12.5cm}
\caption{\label{bate.testcase1} First comparison of the results from 
the protobinary evolution (PBE) code (solid lines) with those from a full SPH 
calculation (dotted lines).  
The graphs show the evolution of a protobinary system 
that was formed in the centre of a collapsing molecular cloud core 
which initially had a uniform-density profile and was in solid-body 
rotation.  The evolution of the (a; upper left) mass ratio, (b; upper 
right) separation, (c; lower) ratio of mass in a circumbinary disc 
to that of the binary are plotted 
as the binary accretes gas from the infalling envelope.}
\end{center}
\end{figure}

To test how accurately the PBE code describes the evolution 
of a `seed' binary as it accretes from its initial to 
its final mass, the PBE results were compared to those from 
full SPH calculations.  Two test cases were performed.  The first 
followed the formation of a binary system from the collapse 
of an initially uniform-density, spherical molecular cloud 
core in solid-body rotation.  The `seed' binary was assumed 
to have a mass ratio of $q=0.6$ and a mass of $1/10$ the
initial cloud mass.  The second test case was similar, except 
that the progenitor cloud was centrally-condensed with a 1/r-density 
distribution and the cloud had a total mass of only 5 times the
`seed' binary's mass.  A full discussion of the test cases
is given by \cite{Bate2000}.

\subsubsection{Test Case 1}

The evolution of the mass ratio, separation and amount of gas
in the circumbinary disc are given for
the PBE code and for a full SPH calculation in Figure 
\ref{bate.testcase1}.  The curves are given as functions of the 
amount of gas that has fallen on to the binary, $M_{\rm acc}$, 
relative to the binary's initial mass. 
In addition, an MPEG animation of the
SPH calculation is included on this CD-ROM as bate2.mpg.
The CPU time required to evolve the SPH
calculation until the entire cloud falls on to the binary in 
is prohibitively long, which, after all, is the reason that 
the PBE code was developed in the first place.  It takes 
$\approx 60$ orbits for the binary to increase its mass by 
a factor of 6 (i.e.~$\approx 60$\% of the total cloud was 
accreted).  The SPH calculation took $\approx 5$ months on 
a 170 MHz Sun Ultra workstation with a GRAvity-PipE (GRAPE) board used
to calculate the gravitational forces and neighbouring SPH particles.  
The evolution with the PBE code took a few seconds!

Although the SPH calculation did not run to completion, 
we can compare the evolution as the binary's mass increases 
by a factor of 6 (Figure \ref{bate.testcase1}).  
Generally, there is good agreement 
between the PBE and SPH codes.  The mass ratio is predicted 
to within 5\% over the entire evolution and the separation 
to within 15\%.  In fact, as discussed in \cite{Bate2000},
the small differences between the PBE and SPH results 
reflect unphysical treatment of the circumstellar discs
by the SPH code rather than a problem with the PBE code.
For example, the slower rate of increase of the mass ratio
initially is due to the circumsecondary disc not being 
resolved correctly in the SPH calculation, and the larger
separation when $M_{\rm acc}\simgreat 3$ is due to 
unphysically-rapid viscous evolution of the 
circumstellar discs which transfers angular momentum into
the binary's orbit too quickly.  The greatest difference between
the PBE and SPH results is that the PBE code predicts that a 
circumbinary disc should be formed around the binary whereas
no circumbinary disc is formed in the SPH calculation.  This is
due to the larger separation of the binary when $M_{\rm acc}\simgreat 3$ 
and the large shear viscosity in the SPH calculation.

\subsubsection{Test Case 2}

Unlike test case 1, the PBE code predicts that a massive 
circumbinary disc should be produced very early in the 
evolution of test case 2.  Thus, test case 2 provides a 
better test of how well the PBE code predicts the formation
of a circumbinary disc and its evolution.  We note that,
although the
PBE code records the amount of gas which settles into a 
circumbinary disc, it does not attempt to take account of the
interaction between the binary and the circumbinary disc.  In
reality, this interaction is expected to result in the 
transfer of angular momentum from the orbit of the binary into
the gas of the circumbinary disc and, hence, in a smaller 
separation.  Furthermore, if the separation decreases, more 
of the infalling gas would be expected to settle into the 
circumbinary disc and, for the same increase in the binary's
mass, the mass ratio should increase more rapidly because the
gas has a greater specific angular momentum relative to that of
the binary.  Thus, if a massive circumbinary disc is formed, 
the PBE code is expected to over-estimate the binary's separation, 
under-estimate the mass in the circumbinary disc, and slightly
under-estimate the mass-ratio of the binary.

\begin{figure}[t]
\begin{center}
\epsfig{file=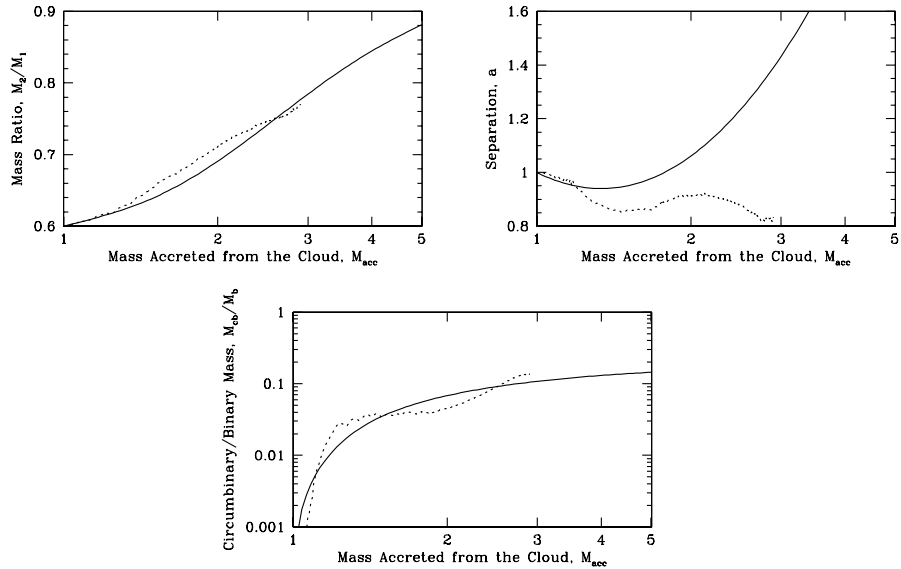, width=12.5cm}
\caption{\label{bate.testcase2} Same as Figure 14, except that the 
protobinary system was formed from a molecular cloud core 
which initially had a density profile of $\rho \propto 1/r$.}
\end{center}
\end{figure}

The evolution of the mass ratio, separation and amount of gas
in the circumbinary disc are given for test case 2 in Figure 
\ref{bate.testcase2}.  An MPEG animation of the
SPH calculation is included on this CD-ROM as bate3.mpg.
To avoid the problems that occurred due to the large shear viscosity
in the SPH calculation for test case 1, the SPH calculation here
uses a formulation with less shear viscosity (see \cite{Bate2000}).
As with test case 1, due to the computational cost, the SPH 
calculations were stopped before all of the gas had fallen on 
to the binary.  The SPH calculation took $\approx 4$ months 
on a 300 MHz Sun Ultra workstation (using a binary tree, 
not a GRAPE board).  During the evolution, the 
binary performed $\approx 40$ orbits and $\approx 60$\%
of the total mass was accreted by the binary or settled 
into a circumbinary disc.

The agreement for the evolution of the mass ratio is even 
better than it was with test case 1 with differences between the PBE
and SPH results of $\simless 3$\%. The separation
follows the prediction of the PBE code to better than $3$\%
until the circumbinary disc begins to form.  Once the circumbinary
disc attains approximately 5\% of the binary's mass, however,
the separation is always smaller than predicted by the PBE code.
As described above, this is expected because the PBE code
neglects the separation-decreasing effect of the interaction 
between the binary and the circumbinary disc.  This also explains
why the PBE code slightly under-estimates the binary's mass ratio.
However, it is pleasing to see that even neglecting the 
interaction between the binary and the circumbinary disc, 
the PBE code still predicts the mass of the circumbinary disc to 
within a factor of 2 of that given by the SPH code during the
entire evolution.

\subsection{The Evolution of Accreting Protobinary Systems}

\begin{figure}[t]
\begin{center}
\epsfig{file=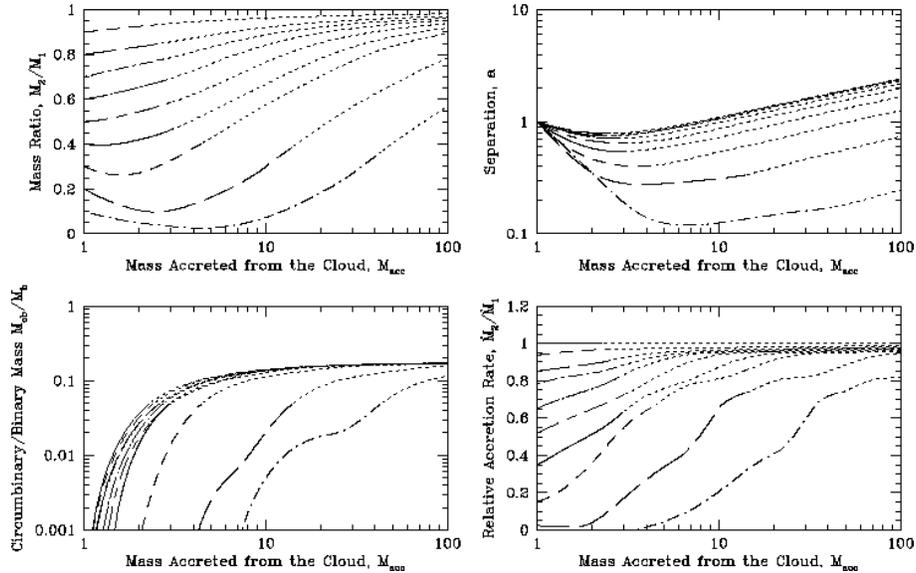, width=12cm}
\vspace{0.5truecm}
\caption{\label{bate.uniform} The evolution of protobinary systems 
formed in the centre of collapsing molecular cloud cores as they 
accrete from the gaseous envelope.  The initial cores have uniform-density 
profiles and are in solid-body rotation.  The evolution of the mass ratio 
(a; upper left), separation (b; upper right), and ratio of the 
circumbinary disc mass to that of the binary (c; lower left), and the 
relative accretion rate (d; lower right) are given as functions of the 
amount of gas that has been accreted from the envelope.  The evolutionary
curves
are given for `seed' binary systems with initial mass ratios of $q=0.1$ to 
$q=1.0$, with various different line types and/or widths for each.  The 
curves are all given by a thin dotted lines once the circumbinary 
disc mass exceeds 5\% of the binary's mass.  Beyond this point, 
the binary's mass ratio and the mass in the circumbinary disc are likely 
to be under-estimated, and the separation is likely to be over-estimated. }
\end{center}
\end{figure}

As we have seen, the PBE code gives a relatively accurate 
description of the evolution of an accreting protobinary, but does
so $\sim 10^6$ times faster than a full hydrodynamic SPH calculation.
This allows us to perform many calculations to 
study how the evolution of a binary as it accretes
to its final mass depends on its initial mass ratio and on the
properties of the molecular cloud core from which it formed.

As examples, we give the evolution of `seed' binaries that form
from two types of molecular cloud core.  A greater range of
molecular cloud cores is considered in \cite{Bate2000}.
Figure \ref{bate.uniform} presents the evolution of
binaries formed from molecular cloud cores which had
uniform-density and were in solid-body rotation before they began
to collapse dynamically.  In Figure \ref{bate.1r}, the 
molecular cloud cores had radial density profiles of $\rho \propto 1/r$
with solid-body rotation, initially.  
Evolutionary curves are provided for `seed' binaries
with initial mass ratios ranging from $q=0.1-1.0$.  

In all cases, the long-term evolution is towards a mass ratio 
of unity, since the material that falls in later has higher 
specific angular momentum relative to that of the binary.  
Thus, the more the binary accretes relative to its initial mass, 
the stronger the tendency is for the mass ratio to be 
driven to unity.  Similarly, the more the binary accretes 
relative to its initial mass, the more likely it is to be surrounded
by a circumbinary disc.  Note that, in the previous sections,
we found that when a massive circumbinary disc is formed, the PBE
code tends to over-estimate the separation, and under-estimate 
the mass of the circumbinary disc and the binary's mass ratio.
Thus, if anything, the evolutionary curves in Figures \ref{bate.uniform}
and \ref{bate.1r} tend to under-estimate the binary's mass ratio and 
the mass of the circumbinary disc.

\subsection{The Properties of Binary Stars}

\begin{figure}[t]
\begin{center}
\epsfig{file=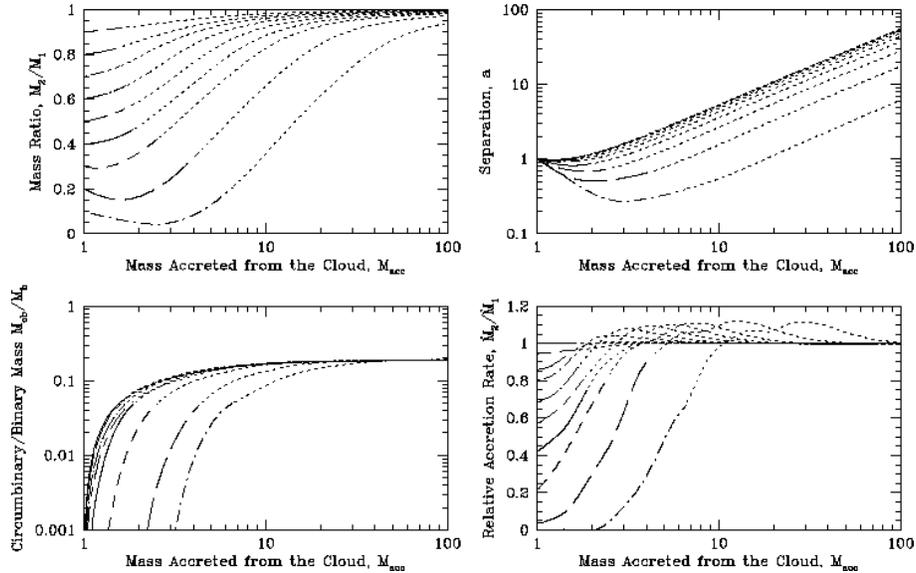, width=12cm}
\vspace{0.5truecm}
\caption{\label{bate.1r} Same as Figure 16, except that the 
protobinary systems were formed from a molecular cloud cores 
which initially had a density profiles of $\rho \propto 1/r$.}
\end{center}
\end{figure}

The aim of developing the PBE code was to make it possible to predict
some of the properties of binary stars and, by comparing
these to the observed properties of binary systems, to constrain 
the initial conditions for binary star formation.  

In order to obtain predictions about the properties of binaries we 
note that, generally, the initial mass of a `seed' binary is smaller 
for those binaries with smaller separations.  This relationship
between a `seed' binary's mass and its separation is observed from
fragmentation calculations \cite{Boss86, BonBat94b} and
is easily understood from a Jeans-mass argument \cite{Bate2000}.  
In order for
fragmentation to occur, the Jeans length at the time of fragmentation
must be less than or approximately equal to the separation of the
binary which is formed.  However, for a constant temperature,
the Jeans mass depends linearly on the Jeans length.  Thus, the
smaller the separation of the `seed' binary, the smaller its initial
mass.  Generally, `seed' binaries with separations $\simless 10$
AU are expected to have masses $\approx 0.01 {\rm M}_{\odot}$, while
for larger separations, the `seed' mass is expected to increase
approximately linearly (i.e. `seed' binaries with separations of
100-1000 AU should have initial masses of 
$\approx 0.1-1.0 {\rm M}_{\odot}$).

This dependence of the initial mass on the separation means that
to form binaries with the same final total mass, the closer systems
need to accrete more material, relative to their initial mass.
Therefore, from the evolutionary curves of Figures \ref{bate.uniform}
and \ref{bate.1r}, closer systems are more likely to have equal-mass
components than wider systems.

This prediction is supported by surveys of main-sequence 
G-dwarf stellar systems.  Duquennoy \& Mayor \cite{DuqMay91} 
found that the mass-ratio distribution, averaged over 
binaries with all separations, increases toward small mass 
ratios.  However, there is mounting evidence that
the mass-ratio distributions differ between short and long-period
systems with the distribution for close binary systems ($P <
3000$ days; $a \simless 5$ AU) consistent with
a uniform distribution \cite{Mazehetal92, HalMayUdr98}.
Thus, relative to wide systems, the close systems are biased toward
mass ratios of unity.

The fraction by which the mass of a `seed' binary must
be increased in order for its mass ratio to approach unity depends
on the conditions in the molecular cloud core.  Generally,
the less centrally-condensed a core is, the easier it is to 
form a binary system with a low mass ratio 
(c.f.~Figures \ref{bate.uniform} and \ref{bate.1r}).
We can use this dependence of the evolutionary curves on the 
type of molecular cloud core to attempt to constrain the initial
conditions for binary star formation.

Duquennoy \& Mayor \cite{DuqMay91} found that binaries containing
G-dwarfs with
separations $\simgreat 30$ AU generally have unequal masses
(typically $q\approx 0.3$).  Such binaries are likely to have
accreted from a few to ten times there initial mass.  For uniform-density
cores (Figure \ref{bate.uniform}), the observed mass-ratio 
distribution can easily be obtained.  Cores with $\rho \propto 1/r$
result in higher mass ratios than uniform-density cores, but it is
still possible to envisage a spectrum of `seed' mass ratios which gives
a final mass-ratio distribution which is consistent with the
observations of wide binaries.

However, close binaries ($\simless 5$ AU) have initial masses
of $\approx 0.01 {\rm M}_{\odot}$.  Thus, they are expected to
have to accrete up to 100 times their initial mass from the infalling
gaseous envelope before systems with G-dwarf primaries are obtained,
yet the observed mass-ratio distribution is approximately flat
(i.e.~approximately 1/2 the binaries have $q<0.5$).  
It is effectively impossible for cores in solid-body rotation 
to produce such a mass ratio distribution if they are 
significantly centrally-condensed (Figure \ref{bate.1r}).  
Even with uniform-density cores most of the `seed' binaries 
would need to have mass ratios $q<0.1$, which is unlikely.

However, the PBE code only evolves circular binaries whereas 
most binaries have significant eccentricity \cite{DuqMay91}.
For the same semi-major axis, eccentric binaries have less 
angular momentum than circular binaries meaning that the 
clouds from which they formed may be rotating more slowly
and, thus, the gas in the envelope may have less angular momentum.
This would result in slower evolution toward equal masses for
eccentric binaries.
Taking the effects of eccentricity into account, it is quite 
possible that the observed binary mass ratios could be 
produced by the collapse of molecular cloud cores
with radial density profiles less centrally-condensed than $\rho 
\propto 1/r$.  However, even accounting for eccentric binaries, 
it seems virtually impossible that the observed G-dwarf binary 
systems could have been formed from molecular cloud cores with
density profiles that were more centrally-condensed 
than $\rho \propto 1/r$.

The above conclusion that closer binaries should have mass ratios
that are biased toward unity compared to wider systems with the
same total mass is just one of many predictions that may be
derived using the PBE code (see \cite{Bate2000}).  
Others include:  closer binaries are 
more likely to have circumbinary discs than wider binaries; 
brown dwarf companions to solar-type stars
should be very rare at separations $\simless 5$ AU, but their
frequency should increase at larger separations.

\section{Conclusions}

The Lagrangian nature of SPH and its inherent ability to
provide finer spatial resolution in regions of higher density 
make it a powerful tool which is ideally suited for 
studying star formation.

Recent advances in the study of binary star formation that 
have been made using SPH include: the realisation that it is
essential that the Jeans mass is always resolved in numerical 
studies of self-gravitating gas;  the ability to perform 
three-dimensional hydrodynamic calculations which follow the 
collapse of a molecular cloud core to stellar densities;
the study of the effects of accretion on a protobinary systems
and the development of a code which enables the evolution of
accreting binary systems to be followed $\sim 10^6$ times 
faster than a full hydrodynamic calculation.  The latter 
of these developments has resulted in the first firm predictions
of the properties of binary stars for one particular model
of binary star formation.


\end{article}

\end{document}